\documentclass[showpacs,10pt,twocolumn,prl,superscriptaddress]{revtex4-1}

\usepackage{amsmath}
\usepackage{amssymb}
\usepackage{graphicx}
\usepackage{amssymb}
\usepackage{graphics}
\usepackage{epsfig}
\usepackage{CJK}
\usepackage{color}

\begin{document}

\def\kfese{K$_{x}$Fe$_{2-y}$Se$_2$~}
\def\kfeses{K$_{0.64(4)}$Fe$_{1.44(4)}$Se$_{2.00(0)}$~}
\def\kfesete{K$_{0.70}$Fe$_{1.46}$Se$_{1.85}$Te$_{0.15}$}
\def\Hc2{$\mu_0H_{c2}(T)$}

\begin{CJK*}{GBK}{}

\title{Nonmetallic Low-Temperature Normal State of K$_{0.70}$Fe$_{1.46}$Se$_{1.85}$Te$_{0.15}$}
\author{Kefeng Wang}
\email{kwang@bnl.gov}
\affiliation{Condensed Matter Physics and Materials Science Department, Brookhaven National Laboratory, Upton, New York 11973, USA}
\author{Hyejin Ryu}
\affiliation{Condensed Matter Physics and Materials Science Department, Brookhaven National Laboratory, Upton, New York 11973, USA}
\affiliation{Department of Physics and Astronomy, Stony Brook University, Stony Brook, New York 11794-3800, USA}
\author{Erik Kampert}
\author{M. Uhlarz}
\affiliation{Hochfeld-Magnetlabor Dresden (HLD), Helmholtz-Zentrum Dresden-Rossendorf, D-01314 Dresden, Germany}
\author{J. Warren}
\affiliation{Instrument Division, Brookhaven National Laboratory, Upton, New York 11973, USA}
\author{J. Wosnitza}
\affiliation{Hochfeld-Magnetlabor Dresden (HLD), Helmholtz-Zentrum Dresden-Rossendorf, D-01314 Dresden, Germany}
\affiliation{Institut f\"{u}r Festk\"{o}rperphysik, TU Dresden, D-01062 Dresden, Germany}
\author{C. Petrovic}
\email{petrovic@bnl.gov}
\affiliation{Condensed Matter Physics and Materials Science Department, Brookhaven National Laboratory, Upton, New York 11973, USA}
\affiliation{Department of Physics and Astronomy, Stony Brook University, Stony Brook, New York 11794-3800, USA}
\date{\today}

\begin{abstract}
The normal-state in-plane resistivity below the zero-field superconducting transition temperature $T_c$ and the upper critical field \Hc2 were measured by suppressing superconductivity in pulsed magnetic fields for \kfesete. The normal-state
resistivity $\rho_{ab}$ is found to increase logarithmically with decrasing temperature as $\frac{T}{T_c}\rightarrow 0$. Similar to granular metals, our results suggest that a superconductor - insulator transition below zero-field T$_{c}$ may be induced in high magnetic fields. This is related to the intrinsic real-space phase-separated states common to all inhomogeneous superconductors.
\end{abstract}

\maketitle
\end{CJK*}

\section{Introduction}

The normal-state properties (including pseudogap at the Fermi surface and proximity of magnetism at the phase diagrams) of unconventional superconductors such as high-$T_c$ copper oxides, iron-pnictides/chalcogenides and heavy-fermions have received much attention~\cite{Scalapino}. The linear temperature dependence of the in-plane resistivity above $T_c$ \cite{linear1,linear2,linear3} is considered to be related to magnetic quantum criticality and antiferromagnetic fluctuations~\cite{linear4}. Since the normal (nonsuperconducting) states of interest appear under extremely high magnetic fields and at low temperatures ($T \rightarrow 0$), only a few studies were performed investigating the normal-state properties of high-$T_c$ superconductors below $T_c$ so far. In copper oxides examples include underdoped La$_{2-x}$Sr$_x$CuO$_4$ and Bi$_2$Sr$_{2-x}$La$_x$CuO$_6$, where $\rho(T)$ diverges logarithmically below $T_c$ and where a superconductor - insulator transition (SIT) is induced in pulsed magnetic fields~\cite{normal1,normal2,normal3,XYShi}. This came as a surprise since insulating state revealed by the magnetic-field induced breakdown of superconductivity was not expected in a conventional metal. Studies (Ba,K)Fe$_2$As$_2$ and LaFeAsO$_{1-x}$F$_{x}$ resistivity suggested a slight upturn below zero-field $T_c$ after the partial suppression of superconductivity up to 61 T pulsed field \cite{BaKFeAs,Kohama}. In SmFeAsO$_{1-x}$F$_{x}$ magnetic-field induced logarithmic insulating behavior in the resistivity spans one decade in temperature above the zero-field T$_{c}$, stems from the large magnetoresistance that extends to temperatures well above the log-T regime and is not linked to the superconducting state \cite{Riggs}.

\kfese exhibits a high-temperature insulator-metal crossover at $T_s \sim 100-200$ K before becoming superconducting at $T_c$ = 30 K~\cite{Guo,Dagotto, HHWen}. Insulating ground states can also be found by tuning parameters such as Fe occupancy or isovalent S substitution on the Se atomic site~\cite{YanYJ,Lei}. As opposed to copper oxides, a dual description including both itinerant and localized carriers/magnetic moments may be required~\cite{mott,YuRong2}. But similar to the copper oxides, it is imperative to probe the normal state below zero-field $T_c$. This is even more so due to the argued granular-like superconductivity arising in nanoscale superconducting grains separated by non-superconducting islands, with a macroscopic superconducting condensate being established via Josephson tunneling among the grains~\cite{sigrist,phase}. In the underdoped cuprates electronic granularity stems from the nature of the chemical bonds; hole-rich superconducting regions may form for low level of hole doping even though the crystal structure is not granular~\cite{phase2}. In \kfese crystals, the phase separation is an intrinsic feature of the crystal structure~\cite{phase3,phase4}.

Here, we measured the normal-state in-plane resistivity below $T_c$ for \kfesete by suppressing superconductivity in pulsed magnetic fields. The normal-state resistivity $\rho_{ab}$ is found to increase logarithmically as $\frac{T}{T_c}\longrightarrow 0$. Similar to granular superconductors and underdoped La$_{2-x}$Sr$_x$CuO$_4$ and Bi$_2$Sr$_{2-x}$La$_x$CuO$_6$ ~\cite{normal1,normal2,normal3,XYShi}, our results suggest for the first time that magnetic-field-driven SIT is induced in high magnetic field in Fe-based superconductors.

\begin{figure}[tbp]
\includegraphics[scale=0.4]{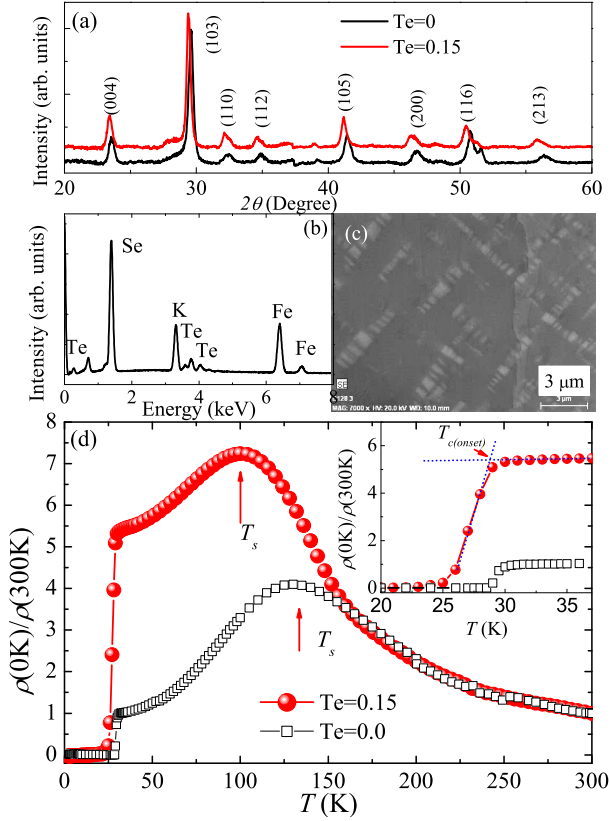}
\caption{(color online) Nominal $7.5\%$ Te substitution of Se in \kfese induces a decrease of the superconducting transition temperature $T_c$ and of the semiconductor-metal transition temperature $T_s$. (a) The powder x-ray diffraction patterns for Te-doped and pure \kfese crystals. The shift of the diffraction peaks toward lower angle shows the expanded lattice in Te-doped crystals. (b) The energy-dispersive X-ray spectroscopy of a typical \kfesete single crystal, which clearly shows the Te peak. (c) The SEM image of a typical single crystal showing stripe phases. (d) Temperature dependence of the resistivity for pure and Te-doped K$_x$Fe$_{2-y}$Se$_2$. The inset shows the magnified part around the superconducting transition temperature. The $T_{c(onset)}$ decreases from 30 to 29.5 K and $T_s$ decreases from 130 to 100 K.
}
\end{figure}

\section{Experimental}

Single crystals of pure \kfese and nominally $10\%$ Te-doped \kfese used in this study were grown and characterized as described previously \cite{Lei}. The powder X-ray diffraction (XRD) spectra were taken with Cu K$\alpha$ radiation ($\lambda = 0.15418$ nm) by a Rigaku Miniflex X-ray machine, confirming phase purity. The cleaved surface image was determined in a JEOL JSM-6500 scanning electron microscope (SEM). The average stoichiometry was determined by energy-dispersive X-ray spectroscopy (EDX) in the same SEM. Four-probe resistivity was
measured in pulsed magnetic fields up to 65 T and in Quantum Design PPMS-14 and PPMS-9. Contacts were attached to the $ab$ plane of crystals in a glove box. Sample
dimensions were measured by an optical microscope Nikon SMZ-800 with 10 $\mu $m resolution. The electric current used in the pulsed field measurement was always 2 mA and the corresponding electric current density in the sample was $\sim 1.39\times10^4$ A/m$^2$.

\section{Results and discussions}

Fig. 1(a) shows the diffraction patterns of both compounds and all reflections can be indexed in the I4/mmm space group. The lattice parameters for \kfesete from the refinement are $a = b = 3.910(3)$ nm and $c = 14.268(6)$ nm, which are larger than for \kfese ($a = b = 3.881(1)$ nm and $c = 14.151(6)$ nm). The expanded crystal lattice implies that Te enters the lattice, which is further confirmed by the EDX large-area elemental analysis which shows the average stoichiometry K$_{0.70(6)}$Fe$_{1.46(3)}$Se$_{1.84(8)}$Te$_{0.15(2)}$ [Fig. 1(b)]. The SEM image [Fig. 1(c)] shows a typical phase-separation pattern. Compared to undoped \kfese \cite{Ryan,WangZ,LiW}, the pattern is inhomogeneous: Stripes can be up to 100 nm large whereas their lower limit is below 50 nm. The local-point analysis on the background (dark) areas gives the stoichiometry K$_{0.78(4)}$Fe$_{1.70(3)}$Se$_{1.84(8)}$Te$_{0.15(2)}$, which is close to the 245 insulating phase \cite{DingX}. About $7.5\%$ Te substitution induces a $T_{c(onset)}$ decrease from $\sim30$ K to $\sim29.5$ K, and metal-semiconductor transition temperature, $T_s$, decrease from $\sim130$ K to $\sim100$ K [Fig. 1(d)].

\begin{figure}[tbp]
\includegraphics[scale=0.4] {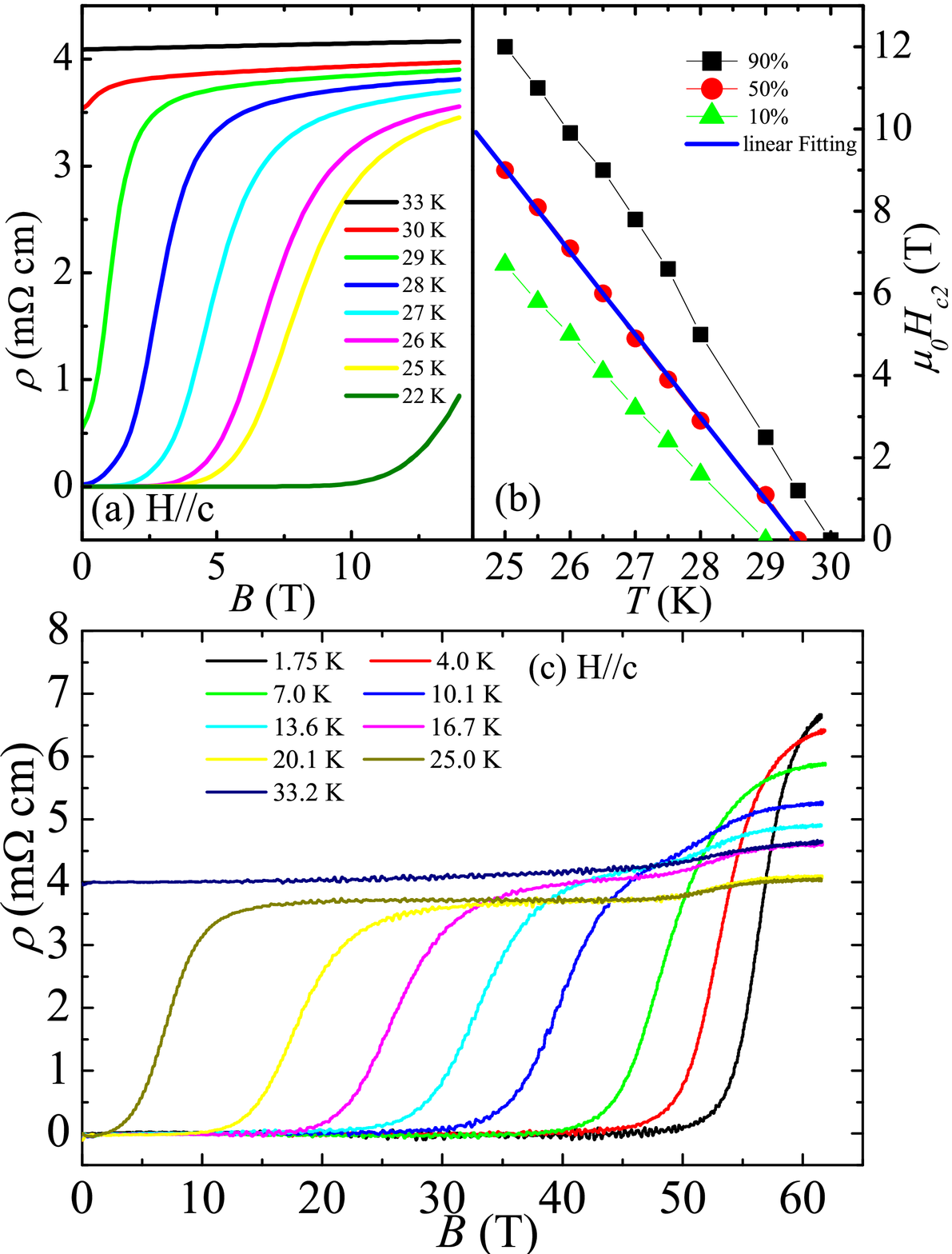}
\caption{(color online) Field dependence of the in-plane resistivity in a \kfesete single crystal with magnetic field applied parallel to the c axis at fixed temperatures. (a) The resistivity data for
temperatures between 22 and $\sim$ 33 K up to 14 T. (b) Temperature dependence of \Hc2 near zero-field $T_{c}$ for the Te-doped crystal with $H\|c$ using different criteria in the resistive transition shown in (a). (c) Resistivity data for temperatures between 1.75 and 33 K in pulsed fields up to 60 T. }
\end{figure}

Near $T_{c}$,  the upper critical field slopes, d\Hc2/d$T|_{T=T_c}$ in \kfese and other ternary/quaternary iron arsenide and selenide superconductors,  are about 5 to 12 TK$^{-1}$ for $H\|ab$
and 1 to 3 TK$^{-1}$ for $H\|c$. Using the Werthamer-Helfand-Hohenberg (WHH) estimate \cite{Werthamer}, results in \Hc2$|_{T=0}$ = (125-275) T and (30-60) T for $H\|ab$ and $H\|c$, respectively \cite{HC21,Mun,Gasparov}.
The upper critical field for $H\|c$ presents almost linear temperature dependence up to 60 T instead of the saturation
predicted by WHH. This is ascribed to multiband effect~\cite{HC21,Mun}. Although the upper critical field for $H\|c$ is larger than the WHH estimate, it is smaller than \Hc2 for $H\|ab$ which is well explained within the WHH theory~\cite{Mun,Gasparov}. For \kfesete, we observe similar slopes of \Hc2 with d\Hc2/d$T|_{T=T_c}\sim 2.1(4)$ TK$^{-1}$ for $H\|c$ [Fig. 2(a-c)] and  d\Hc2/d$T|_{T=T_c}\sim 4.9(3)$ TK$^{-1}$ for $H\|ab$. However, in \kfesete, at lower temperatures \Hc2 is significantly reduced. A magnetic field of 61 T suppresses the superconductivity at 1.7 K for $H\|c$, as opposed to $H\|a$ [Fig. 3(a)] . This is ascribed to a change in the temperature dependence of \Hc2.

\begin{figure}[tbp]
\includegraphics[scale=0.4] {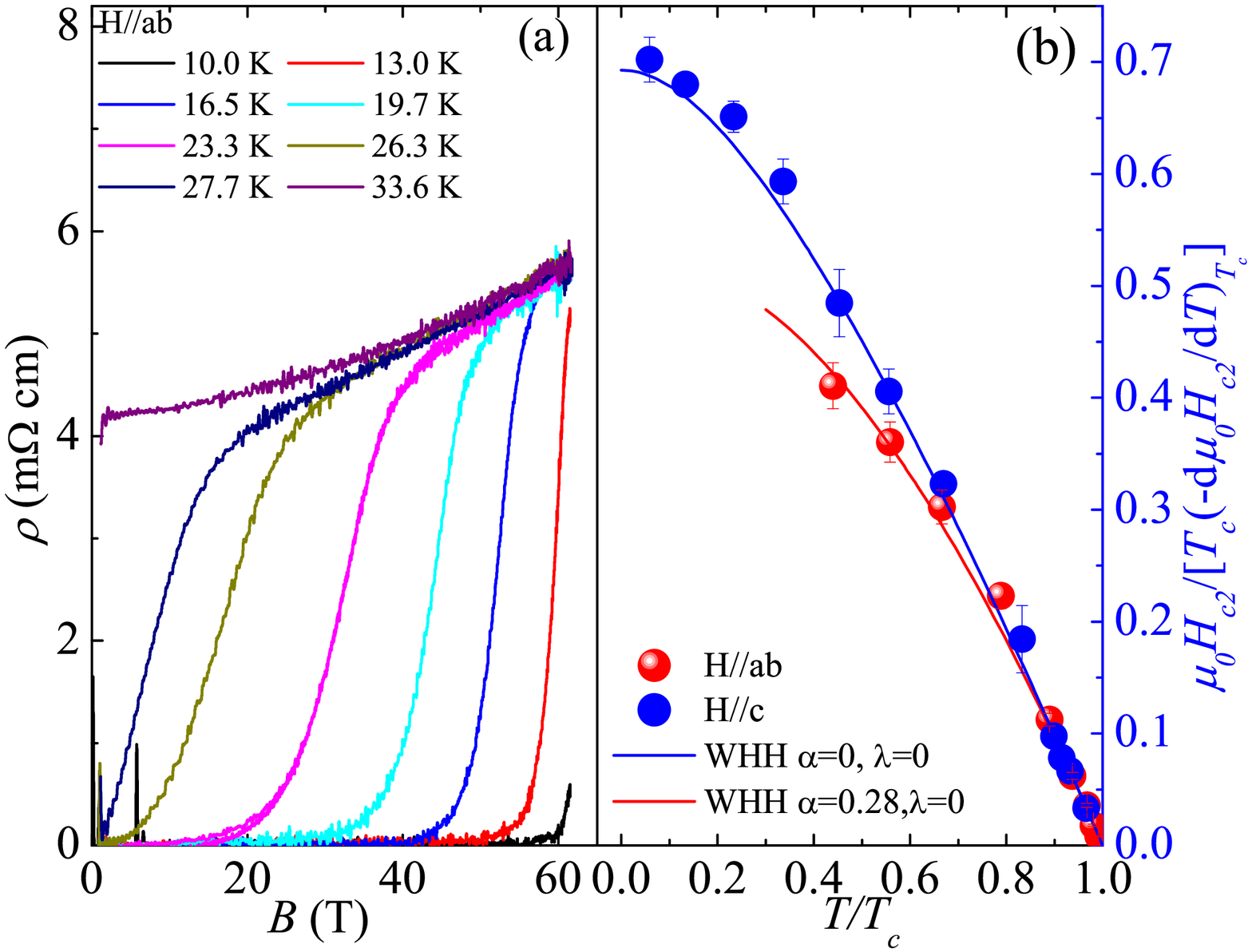}
\caption{(color online) The upper critical fields in \kfesete follow the temperature dependence as predicted by the WHH theory. (a) Field dependence of the in-plane resistivity of \kfesete for pulsed magnetic fields applied perpendicular to the $c$-axis. (b) Temperature dependence of
the upper critical field $H_{c2}(T)$ (plotted as $\frac{\mu_0 H_{c2}}{T_c[-d \mu_0 H_{c2}/dT]_{T_c}}$) for magnetic fields parallel (circles) and perpendicular (balls) to the
$c$ axis, respectively. The solid lines are the fitting results using WHH theory with different parameters.}
\end{figure}

Effects of Pauli spin paramagnetism and spin-orbit scattering can be included in the WHH theory for a single-band s-wave weak-coupling type-II superconductor in the dirty limit
by adding the Maki parameters $\alpha$ and $\lambda_{so}$ \cite{Werthamer,Maki}. Then, \Hc2 is given by
\begin{eqnarray}
\text{ln}(\frac{1}{t})&=&(\frac{1}{2}+\frac{i\lambda_{so}}{4\gamma})\psi(\frac{1}{2}+\frac{h+\lambda/2+i\gamma}{2t})\\ \nonumber
&+&(\frac{1}{2}-\frac{i\lambda_{so}}{4\gamma})\psi(\frac{1}{2}+\frac{h+\lambda/2-i\gamma}{2t})-\psi(\frac{1}{2}),
\end{eqnarray}
where $\psi(x)$ is the digamma function, $\gamma=[(\alpha h)^2-(\lambda_{so}/2)^2]^{1/2}$, and
\begin{eqnarray}
h=\frac{4\mu_0H_{c2}(T)}{\pi^2T_c[-\text{d}\mu_0H_{c2}(T)/\text{d}T]_{T=T_c}}.
\end{eqnarray}
In our data, we observe a convex curvature and a tendency to saturation  for \Hc2 with $H\|c$ at low temperatures [Fig. 3(b)] which can be well understood by the WHH theory with $\alpha=0$
and $\lambda=0$, i.e., with orbitally limited $H_{c2}$: $\mu _{0}H_{c2}^{\ast}(0)$ = -0.693$T_{c}(dH_{c2}/dT)_{T_{c}}$ [the blue curve in Fig. 3(b)]. Hence, the isovalent Te
substitution suppresses the multiband character and promotes a single-band orbital-limited behavior.

The in-plane upper critical field is limited due to spin paragmagnetism since $\alpha=0.28$ and $\lambda_{so}=0$ [red curve in Fig. 3(b)]. When compared to $\alpha_{H\|ab}=5.6$
and $\alpha_{H\|ab}=1.9$ in Tl$_{0.58}$Rb$_{0.42}$Fe$_{1.72}$Se$_2$ and \kfese, respectively \cite{HC21,Gasparov}, the spin-paramagnetic effect in the Te-doped system is very small ($\alpha<1$), but the high anisotropy of \Hc2 for the different principal crystallographic directions remains, suggesting that an anistropic Fermi surface is preserved. Open electron orbits along the $c$ axis render an orbital-limited field unlikely and spin-paramagnetic effects should play an important role for
$H\|ab$. The paramagnetically limited field
$\mu _{0}H_{c2}^{p}(0)$ is given by $\mu_{0}H_{c2}^{p}(0)$=$\mu _{0}H_{c2}^{\ast }(0)$/$\sqrt{1+\alpha ^{2}}$ and $%
\alpha $=$\sqrt{2}H_{c2}^{\ast }(0)$/$H_{p}(0)$, where $\mu _{0}H_{p}(0)$ is
zero-temperature Pauli limited field and the orbitally limited $\mu _{0}H_{c2}^{\ast}$ given above \cite{Maki}. This gives $\mu _{0}H_{c2}^{p,ab}(0)=100$ T whereas $\mu _{0}H_{c2}^{p,c}(0)=45.7$ T. The zero-temperature
coherence length, $\xi(0)$, can be estimated using the Ginzburg-Landau formula $\mu _{0}H_{c2}(0)$ = $\Phi _{0}$/2$\pi \xi^2(0)$, where
$\Phi _{0}$=2.07$\times$10$^{-15}$ Wb, resulting $\xi(0)_{ab}=2.7(1)$ nm and $\xi(0)_{c}=1.8(3)$ nm, respectively.

The reduction of the Maki parameter, $\alpha$, in \kfesete when compared to \kfese cannot be explained by disorder effects since for electronic systems with more disorder one expects a
larger $\alpha$ \cite{Fuchs2}. Therefore, another effect must compete and prevail over the disorder induced by Te substitution. With strong-coupling correction for electron-boson and electron-electron interactions
$\mu _{0}H_{p}(0)\text{=}1.86(1+\lambda )^{\varepsilon }\eta _{\Delta }\eta_{ib}(1-I)$, where strong-coupling intraband corrections for the gap are described by $\eta _{\Delta }$,
$I$ is the Stoner factor $I$=$N(E_{F})J$, $N(E_{F})$ is the electronic density of states (DOS) per spin at the Fermi level, $E_{F}$, $J$ is an effective exchange integral, $\eta _{ib}$
is introduced to describe phenomenologically the effect of the gap anisotropy, $\lambda $ is the electron-boson coupling constant and $\varepsilon $ = 0.5
or 1 \cite{HC27}. The spin-paramagnetic effect can be less pronounced when the Stoner factor becomes small which can be achieved through suppressing either the DOS at the Fermi level or the effective exchange integral $J$~\cite{HC27}. The reduction of both, the multiband character in $H_{c2}(T)$ and the spin paramagnetism when compared to pure
\kfese, implies that electronic states, removed by Te substitution, carry substantial partial density of states at the Fermi level, i.e., that Te substitution reduces the occupancy of heavily
renormalized $d_{xy}$ orbitals \cite{YiM}.

\begin{figure}[tbp]
\includegraphics [scale=0.3]{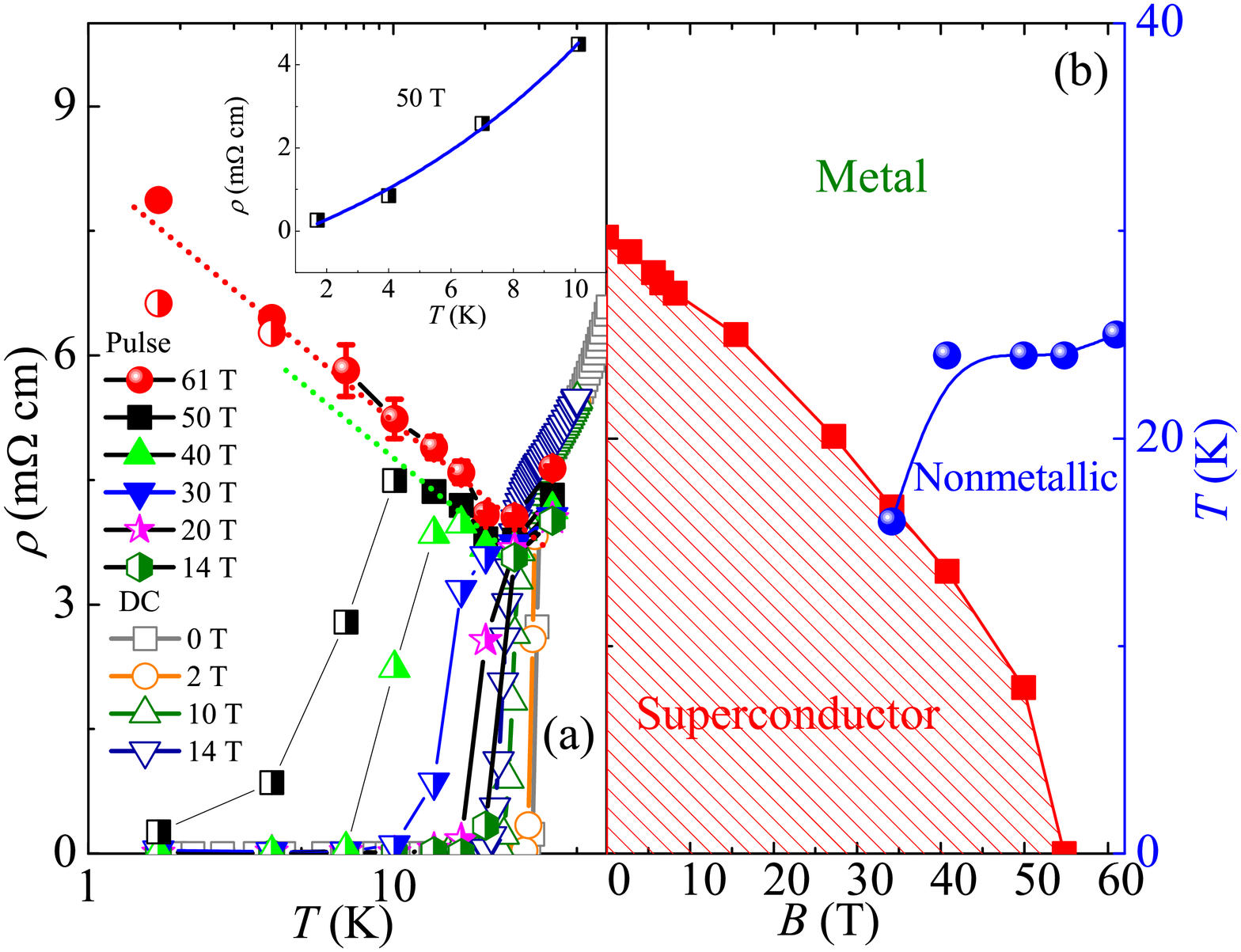}
\caption{(color online) (a) Temperature dependence of the in-plane resistivity in several DC (open symbols) and pulsed (filled symbols)
 magnetic fields for temperature between $1.75$ and $100$ K. Pulsed-field data (filled symbols) near $T_{c}$ is consistent with results in DC and zero fields (open symbols). The two half-filled symbols for 61 T at 1.75 and 4 K denote the extrapolated values of the resistivity after complete suppression of the superconductivity. The
 red and green dashed lines are the linear fitting result for the 61 T data and 50 T data when using logarithmic scale, respectively. Inset: resistivity of the mixed state in 50 T field and the blue dotted line is an exponential fitting result (see text) (b) The
 schematic temperature-field phase diagram of \kfesete.}
\end{figure}

In \kfesete, a magnetic field of 61 T parallel to the $c$ axis completely suppresses superconductivity above
1.75 K [Fig. 2(c) and 3(b)], providing the opportunity to study the normal-state properties as $T\longrightarrow0$. The application of 61 T leads to a negative $d\rho/dT$ resembling
insulating rather than metallic behavior [Fig. 4(a)]. It is apparent that a SIT occurs right below $T_c$ (filled balls) after the superconductivity is completely
suppressed above $H_{c2}$. At 50 T, the resistivity as well shows SIT and is nearly identical to that at 61 T field until superconductivity sets in below about 15 K. We first note that the insulating behavior in high magnetic fields appears only below the zero-field $T_c$. It is also instructive to note that above the zero-field $T_c$ the field dependence of the resistivity up to 61 T is minute for $H\|c$, in contrast to SmFeAsO$_{1-x}$F$_{x}$ \cite{Riggs}. The temperature-magnetic-field phase diagram for
\kfesete is shown in Fig. 4(b).


A SIT was first reported in ultrathin, disordered and granular type-I superconductors \cite{granular,BeloborodovRMP}. Then, a SIT with an insulating resistivity, increasing logarithmically below $T_c$, was reported in high-$T_c$ cuprate superconductors utilizing pulsed magnetic fields \cite{normal1,normal2,normal3}. Such a temperature dependence is different from variable range hopping (VRH) and thermally activated transport observed in semiconductors. Kondo-type magnetic scattering is also unlikely since the spin-flip scattering should be suppressed in 61 T. Evidence for an intrinsic electronic phase separation in the cuprate superconductors \cite{Lang} suggests that the logarithmic increase in the resistivity and the SIT originates from an inhomogeneous state, presumably similar to granular Nb. \cite{BeloborodovRMP,granular2}.

\kfese features a highly defective crystal structure where the magnetic insulating regions are spatially intrinsically separated and coexist with superconducting regions on the
a length scale of 100 nm to 1 $\mu$m \cite{Ryan,WangZ,LiW}. However, as opposed to two-dimensional alternate stacking of insulating and (super)conducting layers, the stripe-type phase separation in \kfese is more three-dimensional \cite{phase,phase2,phase3}. The conductivity in such inhomogeneous samples must include contributions from the insulating ($\sigma_i$) and metallic ($\sigma_m$) regions, i.e., $\sigma=\sigma_{m}+\sigma_{i}$. When $T<T_{c}$, $\sigma_m$ goes to infinity due to superconductivity and the insulating part of the sample is short-circuited ($\sigma=\sigma_{m}=\infty$). But even around $T_{c}$ and in particular in the normal state at high magnetic field when $T\longrightarrow0$, $\sigma\sim\sigma_{m}$ holds as well since insulating regions have a nearly two orders of magnitude higher resistivity \cite{Shoemaker}. This is consistent with the observation that insulating regions
 do not contribute to the spectral weight in angular resolved photoemission data close to $E_{F}$ \cite{YiM}.

Therefore, superconductivity in \kfese is similar to that in granular superconductors, namely a three-dimensional array of superconducting grains in an insulating matrix \cite{BeloborodovRMP,granular2}.
Strong magnetic fields suppress superconductivity in each grain. The possible scenarios explaining a SIT include Anderson localization \cite{disorder} and the bosonic scenario
where the gap is driven to zero by phase fluctuations~\cite{Beloborodov1,Beloborodov2}. The observed logarithmic divergence of the resistivity in the nonmetallic normal state of \kfesete [Fig. 4(a)] does not follow the VRH conduction which is observed for Anderson localization. In the bosonic scenario the Cooper pairs are localized in granules (similar to defects).
For a $H>H_{c2}$, viritual Cooper pairs may form, leading to a reduction of the DOS. When $T\longrightarrow0$, the pairs cannot travel between grains and the resistivity becomes much larger than for a normal metal without electron-electron interactions~\cite{Beloborodov1,Beloborodov2}. In contrast to the spin-paramagnetic critical field, the orbital critical field
depends on the grain size as $H_{c2}^{o}\sim\phi_{0}/R\xi$ where $R$ is the average grain radius and $\xi$ is the coherence length.
Since the upper critical field in \kfesete is about 55(3) T and the grain coherence length is 2.7(1) nm, we estimate a lower limit of the stripe grain size to be R = 21(2) nm.
Near the SIT on superconducting side it is expected that $\rho=\rho_{0}$exp$(T/T_{0})$ \cite{granular2} in high fields due to the destruction of fluctuating quasilocalized Cooper pairs and a concomitant increase of the number of quasiparticles at the Fermi level. Our 50 T resistivity data in the mixed state can be described well by this formula [The blue line in inset of Fig. 4(a)]. Different heat treatment and quenching can bring about changes in the normal-state resistivities, arrangement and connectivity of the superconducting grains \cite{DingX}. This can influence the tunneling conductance in the granular array and, therefore, it would be of great interest to study superconducting fluctuations in such samples in high magnetic fields in the normal state below $T_{c}$ for further comparison with theory \cite{BeloborodovRMP,granular2}.

Even though the logarithmic temperature dependence of the resistivity in copper oxides is consistent with the theory of granular superconductivity if electrons in different CuO planes are incoherent in the low-temperature limit \cite{Beloborodovhitc,Pan}, this is still controversial since the SIT in copper oxides occurs on the metallic side of the Mott limit ($k_{F}l > 1$, where $k_{F}$ is the Fermi wave vector and $l$ is the mean free path) \cite{AndoML}. Approaches offered to explain insulating states in copper-oxide superconductors postulate spin-charge separation~\cite{Emery}, invoke a bipolaron theory of strongly correlated Mott-Hubbard insulators~\cite{Alexandrov}, or propose Mott insulating states in antiferromagnetic vortices within the framework of SO(5) theory of antiferromagnetism and superconductivity~\cite{Demler}. In the light of increasing evidence for insulating states in the different kinds of high-$T_c$ superconductors in pulsed magnetic fields, more theoretical work may be needed to determine which of these approaches is compatible with real-space and/or electronic phase separation in materials of different structure and bonding types.

\section{Conclusion}

In summary, the normal-state in-plane resistivity and the upper critical fields of \kfesete are measured by suppressing superconductivity in pulsed magnetic fields. The temperature dependence of \Hc2 can be well described using the WHH theory, in contrast to \kfese where the linear temperature dependence signals multiband effects. After suppressing superconductivity, the normal-state resistivity, $\rho_{ab}$, is found to be insulating and to diverge logarithmically as $\frac{T}{T_c}\longrightarrow 0$. Similar to the SIT in granular superconductors, our results suggest that the mechanism for the SIT in high magnetic fields is related to intrinsic phase-separated states in this kind of materials.

\section{Acknowledgments}
We thank Myron Strongin and Dragana Popovic for useful discussions. Work at Brookhaven is supported by the U.S. DOE under contract No. DE-AC02-98CH10886. We acknowledge the support of the HLD at HZDR, member of the European Magnet Field Laboratory. CP acknowledges support by the Alexander von Humboldt Foundation.


\end{document}